\documentclass[twoside,twocolumn,9pt]{article}
\usepackage{extsizes}
\usepackage[super,sort&compress,comma]{natbib} 
\usepackage[version=3]{mhchem}
\usepackage[left=1.5cm, right=1.5cm, top=1.785cm, bottom=2.0cm]{geometry}
\usepackage{balance}
\usepackage{mathptmx}
\usepackage{sectsty}
\usepackage{graphicx} 
\usepackage{lastpage}
\usepackage[format=plain,justification=justified,singlelinecheck=false,font={stretch=1.125,small,sf},labelfont=bf,labelsep=space]{caption}
\usepackage{float}
\usepackage{fancyhdr}
\usepackage{fnpos}
\usepackage[english]{babel}
\addto{\captionsenglish}{%
  
}
\usepackage{array}
\usepackage{droidsans}
\usepackage{charter}
\usepackage[T1]{fontenc}
\usepackage[usenames,dvipsnames]{xcolor}
\usepackage{setspace}
\usepackage[compact]{titlesec}
\usepackage{hyperref}

\usepackage{epstopdf}

\definecolor{cream}{RGB}{222,217,201}

\begin{document}

\pagestyle{fancy}
\thispagestyle{plain}
\fancypagestyle{plain}{
\renewcommand{\headrulewidth}{0pt}
}

\makeFNbottom
\makeatletter
\renewcommand\LARGE{\@setfontsize\LARGE{15pt}{17}}
\renewcommand\Large{\@setfontsize\Large{12pt}{14}}
\renewcommand\large{\@setfontsize\large{10pt}{12}}
\renewcommand\footnotesize{\@setfontsize\footnotesize{7pt}{10}}
\makeatother

\renewcommand{\thefootnote}{\fnsymbol{footnote}}
\renewcommand\footnoterule{\vspace*{1pt}%
\color{cream}\hrule width 3.5in height 0.4pt \color{black}\vspace*{5pt}} 
\setcounter{secnumdepth}{5}

\makeatletter 
\renewcommand\@biblabel[1]{#1}            
\renewcommand\@makefntext[1]%
{\noindent\makebox[0pt][r]{\@thefnmark\,}#1}
\makeatother 
\renewcommand{\figurename}{\small{Fig.}~}
\sectionfont{\sffamily\Large}
\subsectionfont{\normalsize}
\subsubsectionfont{\bf}
\setstretch{1.125} 
\setlength{\skip\footins}{0.8cm}
\setlength{\footnotesep}{0.25cm}
\setlength{\jot}{10pt}
\titlespacing*{\section}{0pt}{4pt}{4pt}
\titlespacing*{\subsection}{0pt}{15pt}{1pt}

\fancyfoot{}
\fancyfoot[LO,RE]{\vspace{-7.1pt}\includegraphics[height=9pt]{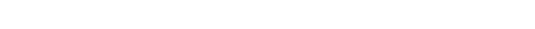}}
\fancyfoot[CO]{\vspace{-7.1pt}\hspace{13.2cm}\includegraphics{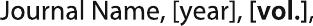}}
\fancyfoot[CE]{\vspace{-7.2pt}\hspace{-14.2cm}\includegraphics{head_foot/RF}}
\fancyfoot[RO]{\footnotesize{\sffamily{1--\pageref{LastPage} ~\textbar  \hspace{2pt}\thepage}}}
\fancyfoot[LE]{\footnotesize{\sffamily{\thepage~\textbar\hspace{3.45cm} 1--\pageref{LastPage}}}}
\fancyhead{}
\renewcommand{\headrulewidth}{0pt} 
\renewcommand{\footrulewidth}{0pt}
\setlength{\arrayrulewidth}{1pt}
\setlength{\columnsep}{6.5mm}
\setlength\bibsep{1pt}

\makeatletter 
\newlength{\figrulesep} 
\setlength{\figrulesep}{0.5\textfloatsep} 

\newcommand{\topfigrule}{\vspace*{-1pt}%
\noindent{\color{cream}\rule[-\figrulesep]{\columnwidth}{1.5pt}} }

\newcommand{\botfigrule}{\vspace*{-2pt}%
\noindent{\color{cream}\rule[\figrulesep]{\columnwidth}{1.5pt}} }

\newcommand{\dblfigrule}{\vspace*{-1pt}%
\noindent{\color{cream}\rule[-\figrulesep]{\textwidth}{1.5pt}} }

\makeatother

\twocolumn[
  \begin{@twocolumnfalse}
{\includegraphics[height=30pt]{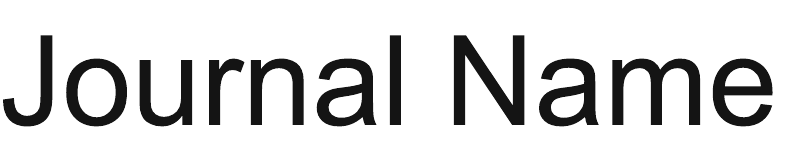}\hfill\raisebox{0pt}[0pt][0pt]{\includegraphics[height=55pt]{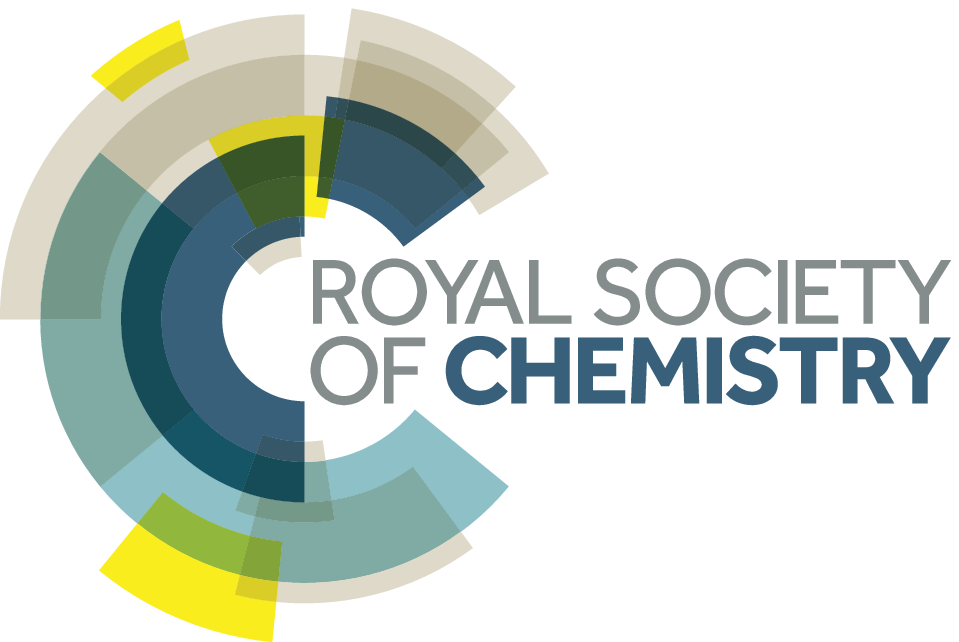}}\\[1ex]
\includegraphics[width=18.5cm]{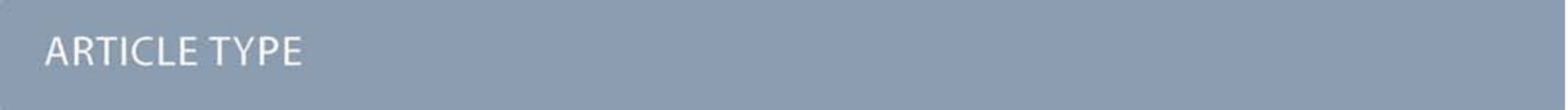}}\par
\vspace{1em}
\sffamily
\begin{tabular}{m{4.5cm} p{13.5cm} }

\includegraphics{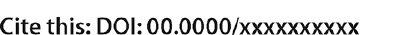} & \noindent\LARGE{Synthetic control over polymorph formation in the d-band semiconductor system \ce{FeS2}} \\
\vspace{0.3cm} & \vspace{0.3cm} \\

 & \noindent\large{KeYuan Ma,$^a$ Robin  Lef\`evre,$^a$ Qingtian Li,$^{b, c, d}$ Jorge Lago,$^{a, e}$ Olivier Blacque,$^{a}$ Wanli Yang,$^{b}$ and Fabian O. von Rohr $^{*,a}$} \\

\includegraphics{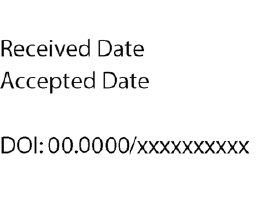} & \noindent\normalsize{Pyrite, also known as fool's gold is the thermodynamic stable polymorph of \ce{FeS2}. It is widely considered as a promising d-band semiconductor for various applications due to its intriguing physical properties. Marcasite is the other naturally occurring polymorph of \ce{FeS2}. Measurements on natural crystals have shown that it has similarly promising electronic, mechanical, and optical properties as pyrite. However, it has been only scarcely investigated so far, because the laboratory-based synthesis of phase-pure samples or high quality marcasite single crystal has been a challenge until now. Here, we report the targeted phase formation via hydrothermal synthesis of marcasite and pyrite. The formation condition and phase purity of the \ce{FeS2} polymorphs are systematically studied in the form of a comprehensive synthesis map. We, furthermore, report on a detailed analysis of marcasite single crystal growth by a space-separated hydrothermal synthesis. We observe that single phase product of marcasite forms only on the surface under the invovlement of \ce{H2S} and sulphur vapor. The availability of high-quality crystals of marcasite allows us to measure the fundamental physical properties, including an allowed direct optical bandgap of 0.76 eV, temperature independent diamagnetism, an electronic transport gap of 0.11 eV, and a room-temperature carrier concentration of 4.14~$\times$~10$^{18}$~cm$^{-3}$. X-ray absorption/emission spectroscopy are employed to measure the band gap of the two \ce{FeS2} phases. We find marcasite has a band gap of 0.73 eV, while pyrite has a  band gap of 0.87 eV. Our results indicate that marcasite -- that is now synthetically available in a straightforward fashion -- is as equally promising as pyrite as candidate for various semiconductor applications based on earth abundant elements.} \\

\end{tabular}

 \end{@twocolumnfalse} \vspace{0.6cm}

  ]

\renewcommand*\rmdefault{bch}\normalfont\upshape
\rmfamily
\section*{}
\vspace{-1cm}


\footnotetext{$^a$ Department of Chemistry, University of Zurich, CH-8057 Z\"urich, Switzerland \\ 
$^b$ Advanced Light Source, Lawrence Berkeley National Laboratory, Berkeley, California 94720, United States \\
$^c$ State Key Laboratory of Functional Materials for Informatics, Chinese Academy of Sciences, Shanghai 200050, China\\
$^d$ University of Chinese Academy of
Sciences, Beijing 100049, China\\
$^e$ Department of Inorganic Chemistry, Univ. del Pais Vasco (UPV-EHU), 48080 Bilbao, Spain\\
$^*$ To whom correspondence shall be addressed: fabian.vonrohr@chem.uzh.ch}




\section{Introduction}
Hydrothermal growth is one of the most important methods for the formation of crystalline solids in both geology and preparative chemistry.\cite{byrappa2015hydrothermal} It involves the precipitation of reactants from a high-temperature aqueous solution at high vapour pressures. This method has been widely applied to synthesize high-quality transition metal chalcogenides crystals. The use of hydrothermal synthesis can be utilised to avoid the formation of high-temperature impurity phases and allow for improved control of the chemical stoichiometry and the formation process.\cite{voiry2015phase,feng2001new,ding2019highly} Pyrite and marcasite are the two naturally occurring iron disulphide \ce{FeS2} polymorphs. Both are known to form in nature under hydrothermal growth conditions.\cite{schmokel2014atomic} Pyrite crystallises in the cubic crystallographic space group \textit{Pa}$\Bar{3}$ with the unit cell composed of an iron face-centred cubic sublattice into which the \ce{S_2^{2-}} ions are embedded.\cite{dodony1996structural} Marcasite on the other hand crystallises in the orthorhombic $Pnnm$ crystal structure. \cite{gronvold1976heat} Both structures have in common that they contain disulphide \ce{S_2^{2-}} ions with short bonding distance, and \ce{FeS6} octahedra as elemental building blocks.\cite{hyde1996marcasite} 

Marcasite is the less commonly occurring natural mineral of the two phases. It is mainly found in natural acidic volcanic hydrothermal vents and sedimentary rocks. \cite{murowchick1986marcasite, craig1993metamorphism} 
This thermodynamically metastable phase transforms to pyrite at high temperature or by exposure to air over long periods of time. \cite{dodony1996structural} The synthesis of the thermodynamic stable polymorph pyrite has been found to be easily achievable in a phase-pure fashion \cite{bouchard1968preparation, wadia2009surfactant,kar2004solvothermal}. However, the synthesis of phase-pure marcasite has proven in the past to be challenging.  There has been various attempts to obtain phase-pure marcasite nanoparticles for battery electrode applications  \cite{fan2017metastable,wu2020rational,voronina2018marcasite,li2015colloidal}. Nevertheless, these reports are not sufficient in order to understand the \ce{FeS2} phase formation. The formation of marcasite occurs under laboratory conditions, as well as in nature, alongside the pyrite formation. This leads to inter-grown and mixed-phase samples. 

Pyrite - consisting solely of non-toxic, earth abundant elements - is widely considered to be a promising candidate for various electronic and catalytic applications. \cite{schmokel2014atomic,jasion2015low,miao2017mesoporous,puthussery2011colloidal,douglas2015ultrafine,ennaoui1986, shukla2016origin,ennaoui1993iron,gao2017pyrite,piontek2019bio} On one hand, pyrite is a promising photovoltaic material due to its strong light absorption and essentially infinite elemental abundance. \cite{shukla2016origin, puthussery2011colloidal} On the other hand, 
pyrite is a very promising material for photoelectrochemical and solid-state Schottky solar cells. \cite{barawi2016hydrogen, voigt2020observation,steinhagen2012pyrite} Furthermore, pyrite has also been found to be a promising catalyst for several electrochemical reactions, e.g. oxygen and hydrogen evolution reactions \cite{tan2019arousing,jasion2015low, miao2017mesoporous} or reduction of \ce{N2} to ammonia \cite{du2020enhanced}.

Previous physical properties measurements on the naturally occurring marcasite have indicated that both polymorphs may have similar electronic and mechanical properties, making also marcasite a promising candidate for various narrow band-gap applications \cite{jagadeesh1980electrical,sanchez2016marcasite}. So far, studies on marcasite were restricted by the unavailability of high-quality and phase-pure samples, especially the preparation of high-quality single crystals under controlled conditions. However, naturally occurring crystals contain inherently impurities, defects, and dopants, making them unsuitable for some physical property measurements. Various efforts have been made for the hydrothermal synthesis of marcasite and pyrite in the past, most commonly resulting in nano-particles.\cite{wadia2009surfactant, fan2017metastable,wu2020rational,voronina2018marcasite,li2015colloidal}. Henceforth, some of the most fundamental properties of marcasite have never been measured on laboratory prepared high purity single crystal samples. 

Here, we report on the straight-forward hydrothermal preparation methods of phase-pure, high-quality marcasite polycrystalline and single-crystalline samples. The individual formation conditions of marcasite and pyrite under acidic conditions are explored in this work, and a predictive and comprehensive synthesis map is presented. We demonstrate experimentally that the pH values of the solution play an crucial role in the phase selective formation process of marcasite and pyrite. We find that high-acidity generally favours the formation of marcasite. However, at extremely low pH values, pH $<$ 0.7 or lower, and at temperatures above 240 $^\circ$C, the formation of phase-pure pyrite samples are observed, which is in contrary to earlier computational predictions \cite{kitchaev2016evaluating}. Subsequently, we perform a detailed experimental measurement and analysis of the physical properties of laboratory-prepared high-purity marcasite single crystal. The measurement results indicate that marcasite is indeed a promising small bandgap semiconductor that has very favourable properties for further deep explorations and other broad potential applications.

\section{Results and Discussion}
\subsection{Structure and composition of hydrothermal polycrystalline \ce{FeS2}} 

\begin{figure}
 \centering
\includegraphics[width= \linewidth]{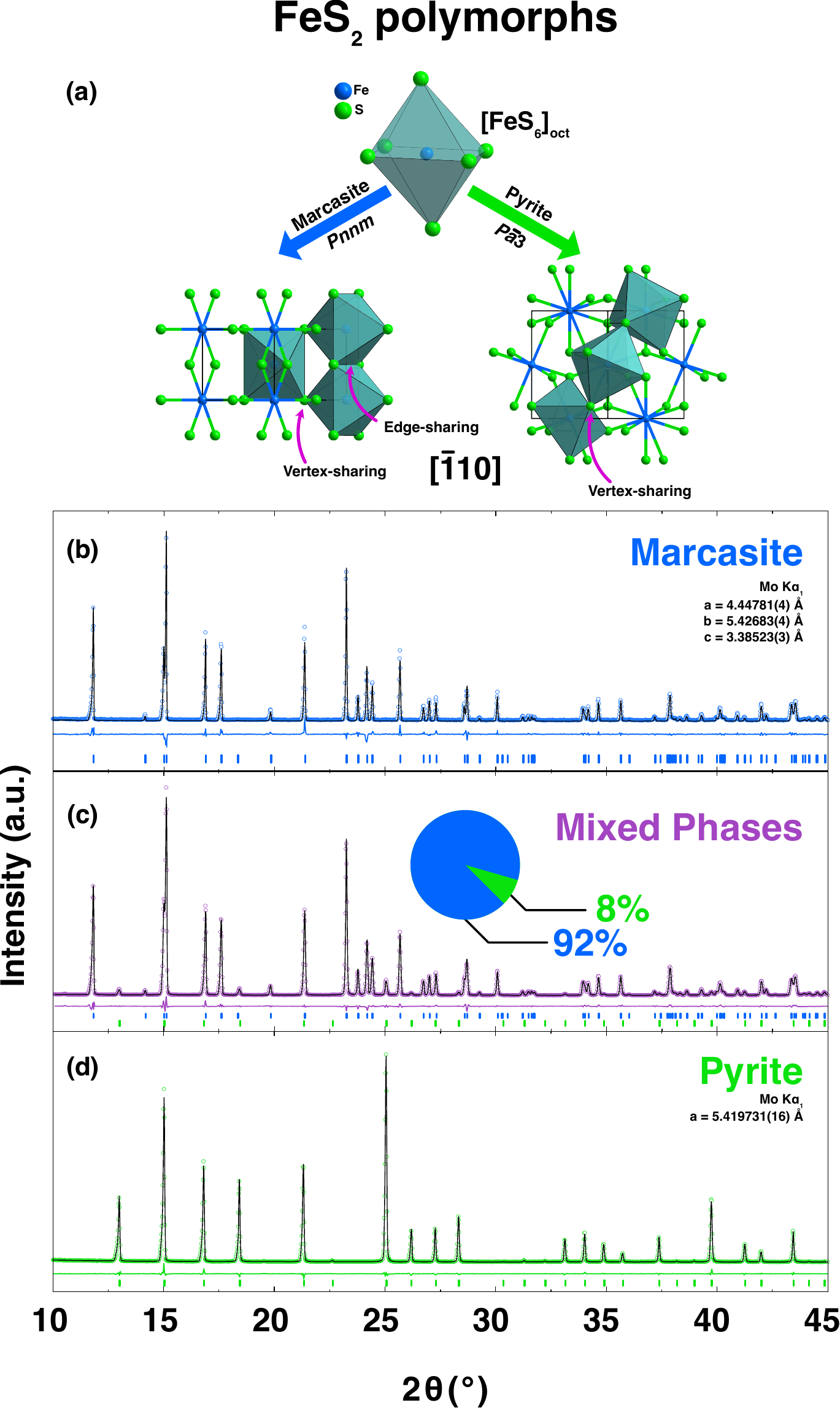}
\caption{(a) Crystal structures along the [-110] direction of the \ce{FeS2} two polymorphs, marcasite and pyrite. PXRD pattern and the corresponding Rietveld refinements of the obtained polycrystalline samples of  phase-pure marcasite (b), a mixture of the two phases (92 \% marcasite) (c), and phase-pure pyrite (d). The coloured circles, black lines, and coloured vertical ticks (marcasite in blue, pyrite in green) stand for observed and calculated patterns, difference between the observed and calculated patterns, and Bragg reflection positions, respectively.}
\label{fig:figure1}
\end{figure}

\ce{FeS2} crystallise in two known structures called polymorphs, pyrite and marcasite. In Figure \ref{fig:figure1}(a), a schematic view of the crystal structures of those two \ce{FeS2} polymorphs is presented. Both crystal structures share the same elemental building blocks, a trigonally distorted \ce{FeS6} octahedra and tetrahedral coordinated sulphur atoms \cite{hyde1996marcasite, schmokel2014atomic}. In marcasite, the octahedra \ce{FeS6} units are edge shared along the unit cell $c$-axis and vertex-linked in $a$, $b$-axis directions by a non-centre collinear \ce{FeS6} octahedra as shown in Figure \ref{fig:figure1}(a). In pyrite, the \ce{FeS6} octahedra are all vertex linked with Fe at the face centred cubic sites. The edge-sharing of the octahedra \ce{FeS6} along the $c$-axis in marcasite causes the Fe–Fe distance to be shorter in this direction than along any other direction in the structure and shorter than the Fe–Fe distance in pyrite \cite{schmokel2014atomic}.   

The experimental synthesis of phase pure marcasite by hydrothermal method is difficult because the two \ce{FeS2} phases often intergrow in solutions. There is still no agreement on how the two \ce{FeS2} phases formed in solutions until now, and previous reports did not give convincing conclusive results due to the complexity of the involved solution reaction systems \cite{murowchick1986marcasite,stanton1991experimental, rickard2007chemistry}. Daniil \textit{et al.} have proposed a quasi-thermodynamic framework for predicting the hydrothermal phase selection synthesis of the two \ce{FeS2} polymorphs, and presented a phase diagram at a wide pH value ranges \cite{kitchaev2016evaluating}. That resulted prediction agrees with previous experimental observations that marcasite is more preferred in acidic solutions \cite{murowchick1986marcasite, rickard2007chemistry}. Guided by this phase diagram, we performed a series of  hydrothermal syntheses under standardised conditions to systematically investigate the experimental formation of marcasite and pyrite over a wide temperature range in the acidic pH value regimes. 

 For this purpose, we reacted elemental sulphur with aquatic \ce{FeCl2} (1 M) solutions under different pH values between 0.6 and 4.0 at varying temperatures between 190 and 250 $^\circ$C for 24 hours. The \ce{FeCl2} has a very good solubility in Millipore water, and the resulting solution has a low pH value to start with. The solid elemental sulphur is insoluble in \ce{FeCl2} solutions, so it does not change the initial pH value of the starting solutions. 

In the series of standardised hydrothermal synthesis reactions, we found four different outcomes for our reactions, depending on the changing pH values and the different synthesis temperatures: (i) there was no resulting precipitate after the reaction (ii) there was a mixture of varying amounts of pyrite and marcasite resulting from the reaction, or (iii) the resulting precipitate was either phase pure pyrite, or (iv) phase pure marcasite. Three representative room temperature PXRD pattern and the corresponding Rietveld refinements of the obtained polycrystalline samples of phase-pure marcasite, pyrite, and a mixture of the two phases are shown in Figure \ref{fig:figure1}(b)-(d). The structural parameters of all the polycrystalline samples were refined from the diffraction data and the results of the refinement parameters are given in supporting information S-Table 1. 

\subsection{Comprehensive synthesis map of marcasite and pyrite \ce{FeS2}}

\begin{figure*}[t]
 \centering
\includegraphics[width= 0.65 \textwidth]{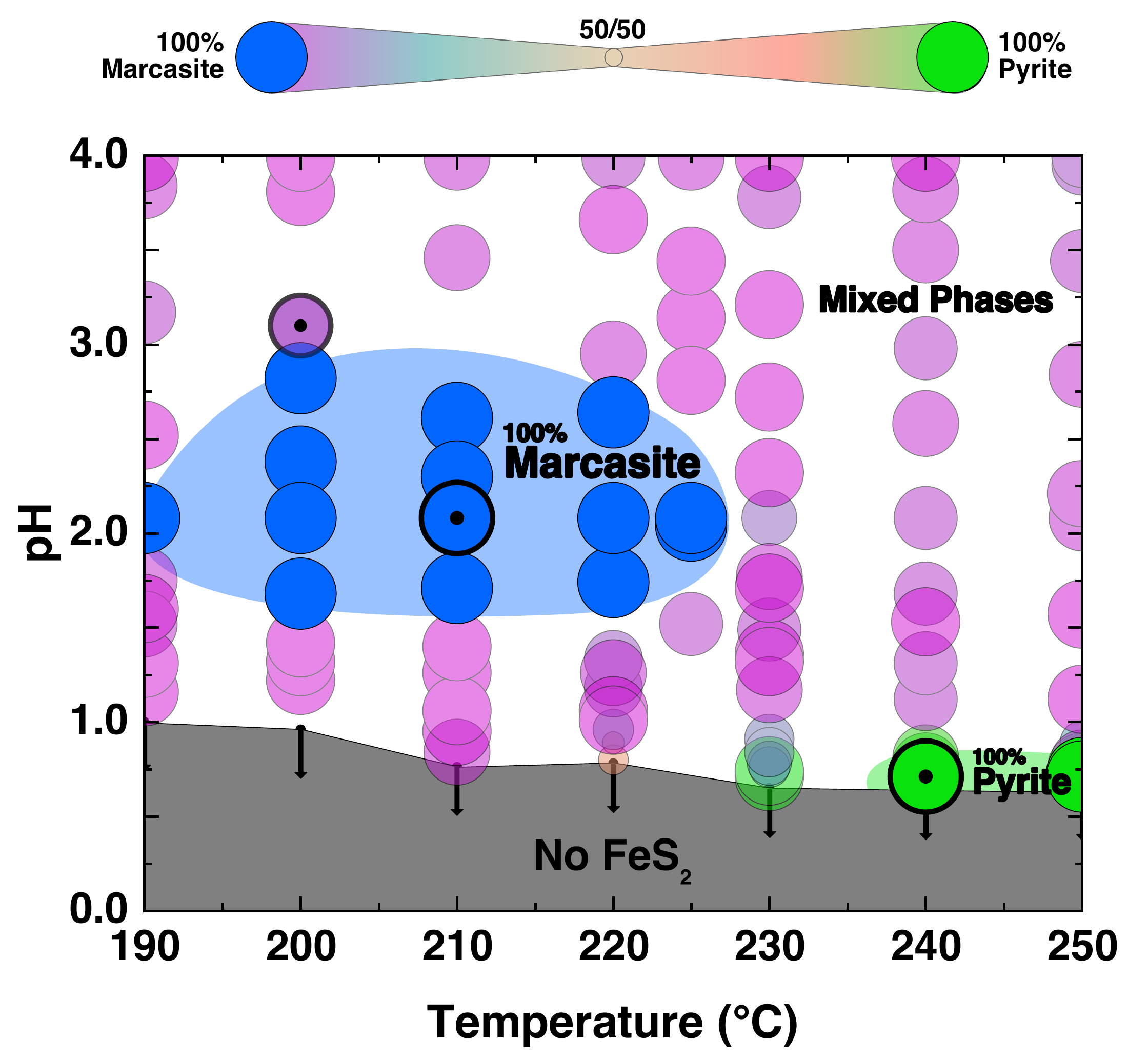}
\caption{Synthesis map of the hydrothermal formation of \ce{FeS2} marcasite and pyrite from a \ce{FeCl2} solution-sulfur reaction system. The synthesis map consists of the four distinct areas: No \ce{FeS2} products are shown in grey, pure \ce{FeS2} marcasite are shown in blue, pure \ce{FeS2} pyrite are shown in bright green and the mixed phase region of consisting of both marcasite and pyrite are shown in gradient colours (marcasite blue, mixtures purple to yellow, pyrite green). }
\label{fig:figure2}
\end{figure*}

 The PXRD refinement results of the obtained products from the performed controlled hydrothermal synthesis(supporting information S-Table 2) allow us to draw a comprehensive temperature–pH synthesis map for the \ce{FeCl2} solution-sulphur reaction system as shown in Figure \ref{fig:figure2}. By comparing this experimentally obtained synthesis map with the phase diagram derived from theoretical predictions, we can have a better understanding of the formation of the two \ce{FeS2} polymorphs.

We find that in the highly acidic region of the synthesis map below pH $ < $ 0.65, neither marcasite nor pyrite forms (grey region). Phase-pure marcasite was obtained in a rather narrow temperature and acidity window: Specifically in a temperature region between 190 and 225 $^\circ$C under acidic conditions of pH values between 1.2 and 2.7 (dark blue region). The formation of phase pure pyrite was found in an extremely narrow temperature and pH window in these acidic hydrothermal conditions. Phase-pure pyrite forms at pH values close to 0.7 and with reaction temperatures of 240 $^\circ$C or more. For most part of the synthesis diagram, we find mixed phase samples, consisting of the two polymorphs.  

The formation of phase pure pyrite and mixed phase samples with pyrite being the dominate phase under these highly acidic conditions was unexpected. High acidic conditions are widely believed to be favourable for the formation of marcasite rather than pyrite \cite{schoonen1991mechanisms}. This has been observed in naturally occurring \ce{FeS2} minerals, and it has been computationally rationalised by the recent first principle calculations based on the surface stability of the two phases as a function of ambient pH within nano-size regimes \cite{kitchaev2016evaluating}.  From the calculation, marcasite is expected to be the dominant phase below a pH value of 4.5. The here developed hydrothermal syntheses' methodology starting from \ce{FeCl2} solutions - which is an acidic salt, giving pH values below a value of 2 to begin with - overall supports the proposed computational model. However, the formation of phase pure pyrite at extremely low pH values, which we observe unequivocally in our experiments, is by this model unaccounted for and is therefore likely due to a different phase formation mechanism. 

\subsection{Growth of \ce{FeS2} marcasite single crystals} 

\begin{figure}
\centering
\includegraphics[width=\linewidth]{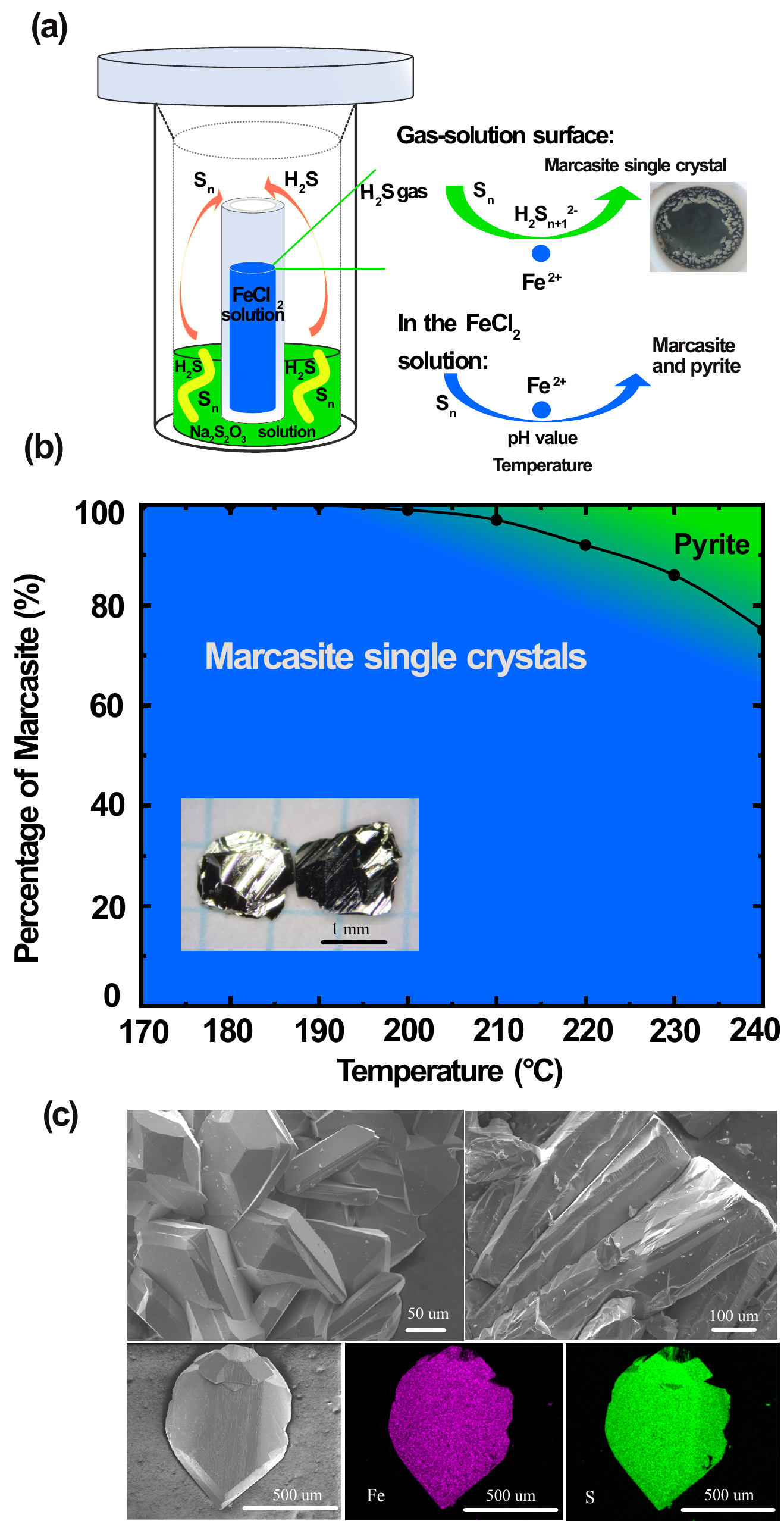}
\caption{(a) Space-separated synthesis of \ce{FeS2} marcasite single crystals. (b) Synthesis map of \ce{FeS2} marcasite single crystal and pyrite from the space-separated synthesis. (c) SEM images and EDS mappings of the obtained marcasite crystals }
\label{fig:figure3}
\end{figure}

Large \ce{FeS2} marcasite single crystals (up to 2 $\times$ 2 $\times$ 1 mm) were prepared by a space-separated hydrothermal synthesis as shown in Figure \ref{fig:figure3}(a). This methodology was first reported by Drabek \textit{et al.} to grow marcasite single crystal at 200 $^\circ$C \cite{drabek2005synthesis}. Here, the space-separated \ce{Na2S2O3} and \ce{FeCl2} solutions were reacted in a temperature range between 170 and 240 $^\circ$C for different durations. Below a reaction temperature of 170 $^\circ$C none or only very little amounts of \ce{FeS2} were formed even after several days of reaction. In the reaction process, \ce{H2S} and S are formed \textit{in situ} from the decomposition of \ce{Na2S2O3}. The \ce{H2S} in the gas phase then reacts in a controlled manner on the surface of the \ce{FeCl2} solution to form marcasite single crystals. The marcasite single crystals were collected after the reaction above the surface of the \ce{FeCl2} solution and on the walls of the inner PTFE beaker. We have observed different morphologies of the resulting marcasite single crystals, depending on the reaction temperature. Below a reaction temperature of 210 $^\circ$C thin plate-shaped crystals were observed (see the photograph in supporting information). At higher temperatures, the marcasite single crystals grow into large independent polyhedrons. At 240 $^\circ$C, marcasite single crystals up to 2 mm were obtained on the walls of the Teflon beaker above the \ce{FeCl2} solution as shown in the photographs in supporting information.

We have also observed the formation of small crystal particles in the \ce{FeCl2} solution as a minority precipitate at reaction temperatures above 200 $^\circ$C. PXRD analysis of the black particles show that these are a mixture of polycrystalline marcasite and pyrite. It is noteworthy that the measured pH values of the \ce{FeCl2} solution after the single-crystal growth reaction was extremely acidic with pH values of about 0.5 (more details see S-Table 3 in supporting information). Hence, these conditions are in agreement with the synthesis map shown in figure \ref{fig:figure2}, where under extremely acidic conditions the formation of pyrite occurs. However, the pyrite impurities were only found in the precipitates collected in the solution, and it cannot be detected in the products collected above the \ce{FeCl2} solution, where the marcasite single crystals are formed. This phenomenon suggests there were different reactions above the \ce{FeCl2} solution surface and in the \ce{FeCl2} solution, which indicate that marcasite and pyrite may form through different reaction pathways. In the computations in reference \citenum{kitchaev2016evaluating} homogeneous processes not involving the gas phase were considered for the hydrothermal synthesis of marcasite and pyrite. However, the heterogeneity and the involvement of the sulphur vapour phases may likely play a crucial role for the formation process under the synthesis conditions presented here. 

Reactions under 180 $^\circ$C for two weeks only yielded very large thin plate-like marcasite single crystals above the \ce{FeCl2} solution (see S-Figure 3(b)in supporting information), without noticeable formation of  black particles in the \ce{FeCl2} solution. Therefore, a space-separated hydrothermal reaction of \ce{Na2S2O3} and \ce{FeCl2} solutions at 170-180 $^\circ$C is the desired condition to get completely phase-pure marcasite single crystals.  

The reaction formation mechanism of marcasite and pyrite is still not clear until now \cite{rickard2007chemistry}. Our results seem to be in good agreement with the previous observation that natural marcasite is formed in the presence of disulphide species and  protonated \ce{S_{n}$^{2-}$} species \cite{murowchick1986marcasite, stanton1991experimental, rickard2007chemistry}. If the reaction temperature is high enough, we speculate that a small fraction of the elemental sulphur will go into the \ce{FeCl2} solution and react there. This reaction pathway may likely be responsible for the formation of polycrystalline marcasite and pyrite precipitates in the solution at reaction temperatures above 200 $^\circ$C, in agreement with our synthesis map discussed above. Increasing the reaction temperature therefore increases the likeliness of this process to occur. In figure \ref{fig:figure3}(b), we present a summary of the temperature-dependant phase formation between \ce{FeS2} marcasite and pyrite by a space-separated reaction growth technique. In this context, it is especially noteworthy that not space-separated reactions of \ce{Na2S3O3} and \ce{FeCl2} solutions, at any conditions leads to the formation of mixtures of marcasite and pyrite and never to phase pure products.

\begin{figure*}[t]
 \centering
\includegraphics[width= 0.7\textwidth]{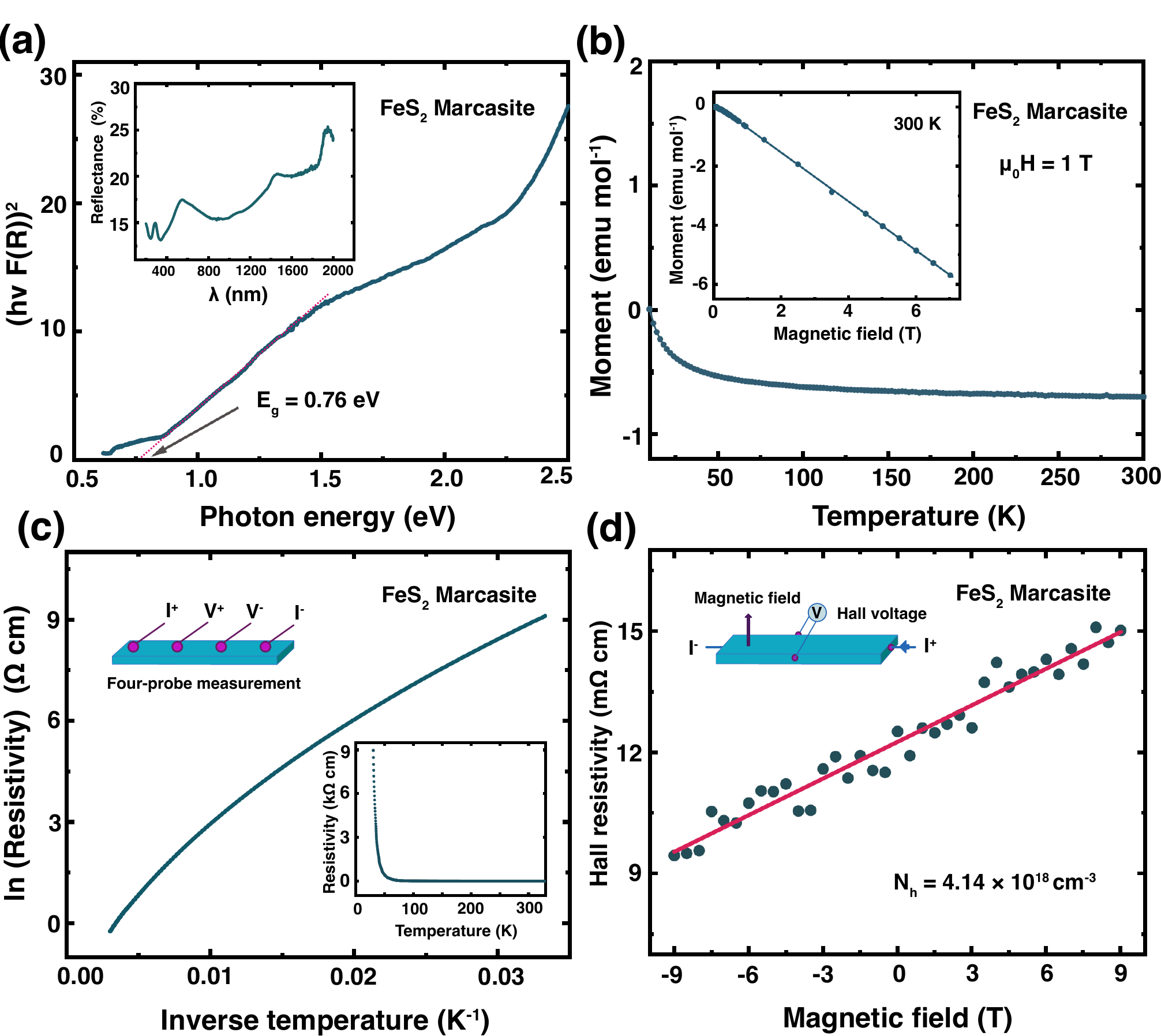}
\caption{(a) The direct allowed optical band gap determination of the marcasite single crystal (inset UV-Vis-NIR spectrum). (b) Temperature dependent magnetic susceptibility of marcasite single crystal in a temperature range of $T$ = 10 K to 300 K, measured in a magnetic field of $\mu_0 H$ = 1 T. (c) Electrical transport measurements in marcasite single crystals in a temperature range of $T$ = 10 K to 300 K. (d) Hall measurements on marcasite single crystals measured at room temperature in magnetic fields between $\mu_0 H$ -9 T and 9 T.}
\label{fig:figure4}
\end{figure*}

In figure \ref{fig:figure3}(c), we show the SEM images of the polyhedron micro-morphology of marcasite. The composition of the \ce{FeS2} marcasite crystal was investigated using energy-dispersive X-ray spectroscopy (EDS) analysis. The corresponding element mapping images show the homogeneous nature of the as-prepared material, in which Fe and S are uniformly distributed in the mapping image of \ce{FeS2} marcasite. EDS investigations of both marcasite single crystals grown at low and high temperatures led to no evidence of undesired elements, and the atom ratio of Fe/S was almost perfectly determined to be 1:2 (Supporting Information S-Figure 6, 7).  We also confirmed the atomic ratio of Fe/S by thermogravimetric analysis in \ce{O2} atmosphere, which also gave an almost perfect 1:2 atomic ratio (Supporting Information S-Figure 8). This combined composition analysis results indicate that the obtained marcasite single crystal are of high quality without sulphur vacancies, in contrary to the naturally occurring marcasite, which usually has substantial sulphur vacancies \cite{sanchez2016marcasite}. 

The Raman spectra of the \ce{FeS2} marcasite shows two notable vibrations bands at 322 and 384 cm$^{-1}$, which are different from pyrite (supporting information S-Figure 9). The obtained large marcasite single crystal from our hydrothermal synthesis can be converted to pyrite single crystal by annealing at 450 $^{\circ}$C under argon atmosphere  (Supporting Information S-Table 4).  
The availability of high-quality single crystals enables us to perform single crystal X-ray diffraction measurements on laboratory grown marcasite \ce{FeS2} single crystals. The details of the analysis are summarized in Table \ref{table:Table1}.

\begin{table}
 \caption{Crystallographic data for single-crystal \ce{FeS2} marcasite}
 \begin{tabular}{c c} 
 \hline 
 Formula & \ce{FeS2} \\ 
 Formula weight ($g \, mol^{-1}$) & 119.97 \\
 Temperature ($K$) & 160(1) \\ 
 Crystal system & orthorhombic \\
 Space group & $Pnnm$ (58) \\
 $a$ (\AA) & 4.4474(4) \\
 $b$ (\AA) & 5.4287(4) \\
 $c$ (\AA) & 3.3852(3) \\
 $\alpha$, $\beta$, $\gamma$ ($^{\circ}$) & 90 \\
V (\AA$^3$) & 81.731(12) \\
Z & 2 \\
Calculated density ($g \, cm^{-3}$)& 4.875 \\
 $\mu$ (mm$^{-1}$) & 11.144 \\
 F (000) & 116.0 \\
 Radiation & Mo K$\alpha$ (0.71073 \AA)\\
 Crystal size ($mm^3$) & 0.09 $\times$ 0.06 $\times$ 0.015 \\
 Recording range 2$\theta$ ($^{\circ}$) & 11.858 --- 55.732\\
 Index ranges & -5 $\leq$ h $\leq$ 5, \\ 
 & -7 $\leq$ k $\leq$ 7, \\ 
 & -4 $\leq$ l $\leq$ 4 \\  
 Reflections collected & 979 \\ 
 Independent reflections &  112 [R$_{int}$ = 0.0582, \\ 
 & R$_{sigma}$ = 0.0244 ]    \\
 Data/restraints/parameters & 112/0/12 \\
 Goodness-of-fit on F$^2$ & 1.187  \\
 Final R indexes [I $\geq$ 2$\sigma$ (I)] & R$_1$ = 0.0253, wR$_2$ = 0.0631 \\ 
 Final R indexes [all data] & R$_1$ = 0.0266, wR$_2$ = 0.0638 \\  
 Largest diff. peak/hole/ e \AA$^{-3}$  &
1.84/-0.93\\
   \hline
\end{tabular}
\centering
\label{table:Table1}
\end{table}

\subsection{Physical properties of laboratory prepared  \ce{FeS2} marcasite single crystals} 

Previously, the physical properties of marcasite were primarily investigated by measurement on natural mineral single crystal samples. However, those samples have substantial structure deficits (e.g. sulphur vacancies) and other impurities \cite{sanchez2016marcasite,jagadeesh1980electrical,garg1991magnetic}. Here, based on the laboratory prepared high-quality marcasite single crystals, we systematically measured its fundamental physical properties and compared them with previous measurement results on natural mineral samples.

We have measured the optical band gap of marcasite by means of UV-Vis-NIR diffuse reflectance spectrum in the region of 200-2000 nm presented in the inset of figure \ref{fig:figure4}(a). The band gap value was obtained using the Tauc Plot method with the equation: ($\alpha$h$\nu$)$^n$ = A(h$\nu$ - E), where h$\nu$ is photon energy, $\alpha$ is absorption coefficient, A is a proportional constant and E is band gap \cite{sanchez2016marcasite}. The absorption coefficient $\alpha$ is replaced by F(R), which is calculated from the acquired diffuse reflectance spectrum with the Kubelka–Munk function: F(R) = {(1-R)$^2$}/{2R} \cite{sanchez2016marcasite}. The obtained data were plotted to: h$\nu$ - (h$\nu$ F(R))$^n$, with n = 2.0 for direct allowed transition and n = 0.5 for indirect allowed transition. The intercept of the tangent linear extrapolation from the curve line on the energy axis is interpreted as the optical band gap. In figure \ref{fig:figure4}(a), a photon energy corresponding to the direct allowed optical absorption gap of $E \approx$ 0.76 eV is clearly defined, and for the indirect allowed gap analysis,  we got $E \approx$ 0.64 eV (supporting information S-Figure 10). These values are slightly lower than those measured on natural marcasite minerals, which gave an indirect allowed transition gap of 0.83 eV \cite{sanchez2016marcasite}. For a direct comparison, we also performed the same measurement on pyrite single crystals obtained by annealing the marcasite crystals. Here, we obtained a direct allowed optical band gap of $E \approx$ 0.84 eV and an indirect allowed optical band gap of $E \approx$ 0.65 eV for pyrite (supporting information S-Figure 10).

In Figure \ref{fig:figure4}(b) we show the temperature dependent magnetic susceptibility $M$($T$) between $T =$ 10 and 300 K in an external magnetic field of $\mu_0 H =$ 1 T. The inset shows the field-dependent magnetic measurements $M$(H) at temperatures of $T =$ 300 K. The magnetic susceptibility of marcasite single crystal is found to be nearly temperature independent and diamagnetic, which is in good agreement with theoretical predictions but has never been experimentally shown on previous investigations \cite{garg1991magnetic}. At very low temperature $T <$ 10 K, we find a so-called Curie-tale, which is caused by tiny magnetic impurities, as expected for any material. The iron in marcasite is in the oxidation state Fe(II) in a d$^6$ configuration, hence the magnetic ground-state is a low-spin state with all the $t_{2g}$ orbitals fully occupied and the $e_g$ orbitals empty. This results in a spin state of $S$ = 0 and overall diamagnetic properties \cite{li2015effect}, as observed in the presented measurement. Henceforth, phase-pure marcasite displays the same magnetic configuration as phase-pure pyrite \cite{burgardt1977magnetic}. This results indicate that the chosen reaction path for obtaining marcasite results in remarkably high-quality samples without notable magnetic impurities.

In the inset of figure \ref{fig:figure4}(c), we present the temperature dependence of the electrical resistivity $\rho$(T) of phase pure \ce{FeS2} marcasite measured using the four-probe technique. The resistivity increases with decreasing temperature as expected for a semiconductor. The ln($\rho$) versus 1/T in figure \ref{fig:figure4}(c) is not a linear Arrhenius behaviour in the measured temperature range, indicating thermal activation is not the only factor that determines the electrical transport \cite{li2015effect}. The high-temperature linear region (280 -330 K) in the ln($\rho$) versus 1/T can be attributed to the intrinsic region of $\rho$ and can be expressed by $\rho = \rho_0 exp(E_a/2k_B T)$, where $\rho_0$ is a pre-exponential constant, $\rm k_B$ represent the Boltzmann constant, and $\rm E_a$ is the activation energy (supporting information S-Figure 11(a)). The fitting of our data results in an intrinsic transport gap of $\rm E_a$ of 0.11 eV, which is smaller than the previous reported 0.38 eV measured on natural marcasite mineral \cite{jagadeesh1980electrical}. The difference in the transport band gap may likely be caused by doping or sulphur deficiency in natural marcasite. 

We have performed Hall measurement on the marcasite single crystal to determine the charge carrier type and density. The Hall resistivity $\rho_{\rm H}$ measurement at room-temperature is presented in figure \ref{fig:figure4}(d). Based on the fitted line, we obtain a Hall coefficient of $R_H =$ 1.51 $\times$ 10$^{-6}$ C$^{-1}$ m$^{3}$ for marcasite (supporting information S-Figure 11(b)). The positive Hall coefficient indicate that marcasite is a p-type semiconductor, and the carrier concentration ($N_h$) is calculated to be 4.14 $\times$ 10$^{18}$ cm$^{-3}$ according to $N_h = 1/eR_H$, where e~=~1.6~$\times$~10$^{-19}$ C is the elementary charge. The above measurement results suggest  marcasite is a d-band semiconductor, with similar properties as its polymorphic form pyrite, making it a promising material for semiconductor applications.

\subsection{Band gap measurement of \ce{FeS2} marcasite and pyrite single crystals}

The information of electronic structure and band gap are very important for us to understand the properties of marcasite. Nevertheless, the reported band gap values of marcasite in previous works are confusing and controversial due to the used different measurement methods on the natural mineral samples.\cite{jagadeesh1980electrical,sanchez2016marcasite} Meanwhile, some related first principle theoretical calculations yielded very different values based on different assumption models\cite{sun2011first,dzade2017periodic}. Hence, there is still no agreement on the band gap value of marcasite until now. Theoretical calculations show the $E_F$ of both marcasite and pyrite are mainly
S 3p character hybridised with Fe 3d ($t_{2g}$) States \cite{schena2013first}. Soft X-ray spectroscopy at the S L-edge can be a reliable approach to obtain the band gaps in \ce{FeS2} \cite{wadia2009surfactant}. To elucidate the electronic structure difference between the two \ce{FeS2} phases, here we performed X-ray absorption and emission spectroscopy (XAS and XES) measurements on the obtained marcasite and pyrite single crystals at the S L-edge.

\begin{figure}
 \centering
\includegraphics[width= \linewidth]{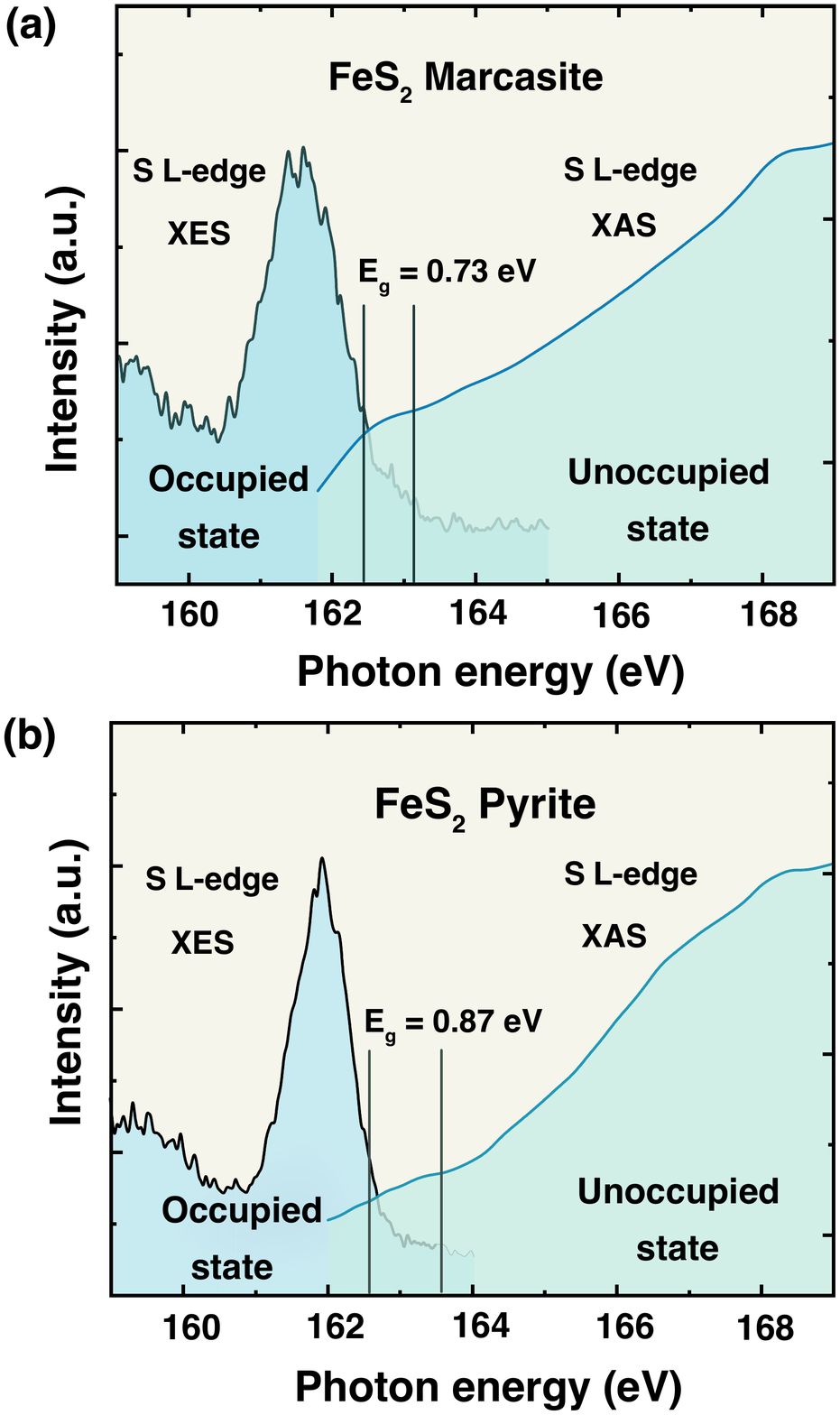}
\caption{ Top of the valence band and bottom of the conduction band indicating the band gap of  \ce{FeS2} marcasite and pyrite.}
\label{fig:figure5}
\end{figure}

We have obtained the XES spectrum of marcasite and pyrite single crystals under different excitation photon energies from 163.4 eV to 198.2 eV (Supporting Information S-Figure 12). We find that the XES spectra of marcasite and pyrite are similar in shape and structure. For both materials, the intensive band around 145-153 eV arises predominantly from levels with 3s character, while the upper valence band being mostly of 3p character shows a much weaker intensity due to the dipole selection rule of the XES process \cite{wadia2009surfactant}. The relatively narrow features at the high emission energy of around 162 eV are attributed to S 3p states that are hybridised with Fe 3d states.  To determine the band gap from the valence band maximum to the conduction band minimum, we plot the S L-edge XES spectrum with excitation energy of  168.6 eV together with S 2p XAS spectrum. Here, we used the method proposed by Gilbert \cite{gilbert2009band} to determine the threshold values of valence-band maximum and conduction-band minimum. In the supporting information S-Figure 13, we showed the fitting process by taking the derivative to locate the inflection points in the valence-band and conduction-band onsets. Based on this analysis, the band gap for marcasite is determined to be 0.73 eV and pyrite is 0.87 eV as indicated in the figure \ref{fig:figure5}(a) and (b), respectively. These values are very close to the band gap values calculated by UV-Vis-NIR diffuse reflectance spectrum for marcasite and pyrite.

\section{Conclusion}

We have presented a pathway to obtain high-quality polycrystalline and single-crystalline samples of \ce{FeS2} by hydrothermal synthesis. We have established a pH and temperature dependent synthesis map for the formation of marcasite and pyrite in highly acidic solutions. We found that phase-pure marcasite is formed in a narrow acidity pH = 1.5-2.7 and temperature region of $T$ = 190-225 $^\circ$C. We confirmed theoretical findings and geological observations that marcasite is preferably formed under acidic conditions. We, however, found that under extremely acidic conditions pH $<$ 0.7 and temperatures above 240 $^\circ$C phase-pure pyrite is formed, which is in contrary to theoretical predictions. By applying a space-separated hydrothermal synthesis, we were able to synthesise large marcasite single-crystals, and to study their physical properties. We find the samples obtained by this method to be of remarkably high-quality, showing e.g. a temperature-independent diamagnetic behaviour, suggesting extremely few defects and vacancies. We, furthermore, find that laboratory grown \ce{FeS2} marcasite is a semiconductor with an optical bandgap of $E =$ 0.76 eV and an electronic transport gap of $E_g =$ 0.11 eV. Soft X-ray spectroscopy measurement on marcasite single crystals shows a band gap of 0.73 eV. These values are comparable to those of pyrite, making \ce{FeS2} marcasite a promising alternative to the already widely studied pyrite, which has been proposed to be electronically most promising for the realisation of solar cells and photo-electrochemical cells. The here presented directly targeted synthesis approach for \ce{FeS2} marcasite may therefore spark efforts to establish this material for various semiconductor applications.

\section*{Author contributions}
FvR designed the experiment. KM synthesised the materials. KM, RL, JL, QL, OB, WL, and FvR performed the measurements and did the data analysis. KM and FvR wrote the manuscript with comments from all the authors.

\section*{Acknowledgements}
This  work  was  supported  by  the  Swiss National Science Foundation under Grant No. PCEFP2\_194183. X-ray spectroscopy data were collected at the Advanced Light Source, a U.S. DOE Office of Science User Facility under contract no. DE-AC02-05CH11231. The PXRD data and the refinements are being made publicly available at https://doi.org/10.6084/m9.figshare.14529057.v1.



\balance


\bibliography{rsc} 
\bibliographystyle{rsc} 

\end{document}